\documentclass[useAMS,usenatbib]{mn2e}

\usepackage{graphics}

\title[A jet ejection of the ULX Holmberg~II~X-1 ]{The evolution of a jet ejection of the ultraluminous X-ray source Holmberg~II~X-1}
\author[D. Cseh et al.]{D. Cseh$^{1}$\thanks{E-mail:
d.cseh@astro.ru.nl}, J. C. A. Miller-Jones$^{2}$, P. G. Jonker$^{3,1}$, F. Gris\'{e}$^{4}$, Z. Paragi$^{5}$, S. Corbel$^{6,7}$, 
\newauthor H. Falcke$^{1}$, S. Frey$^{8}$, P. Kaaret$^{9}$, E. K\"{o}rding$^{1}$
\\
$^{1}$Department of Astrophysics/IMAPP, Radboud University Nijmegen, P.O. Box 9010, 6500 GL Nijmegen, The Netherlands\\
$^{2}$International Centre for Radio Astronomy Research - Curtin University, GPO Box U1987, Perth, WA 6845, Australia\\
$^{3}$SRON, Netherlands Institute for Space Research, Sorbonnelaan 2, 3584 CA Utrecht, The Netherlands\\
$^{4}$Observatoire astronomique de Strasbourg, Universit\'e de Strasbourg, CNRS, UMR 7550, 11 rue de l'Universit\'e, F-67000 Strasbourg, France\\
$^{5}$Joint Institute for VLBI in Europe, Postbus 2, 7990 AA Dwingeloo, The Netherlands\\
$^{6}$Laboratoire AIM (CEA/IRFU-CNRS/INSU-Universit\'{e} Paris Diderot), CEA/DSM/IRFU/SAp, F-91191 Gif-sur-Yvette, France\\
$^{7}$Station de Radioastronomie de Nan\c{c}ay, Observatoire de Paris, CNRS/INSU, USR 704 - Univ. Orl\'eans, OSUC, 18330 Nan\c{c}ay, France\\
$^{8}$F\"{O}MI Satellite Geodetic Observatory, P.O. Box 585, H-1592 Budapest, Hungary\\
$^{9}$Department of Physics and Astronomy, University of Iowa, Iowa City, 52240, USA\\
}

\begin{document}
\date{Draft}
\pagerange{\pageref{firstpage}--\pageref{lastpage}} \pubyear{2015}
\maketitle
\label{firstpage}

\begin{abstract}
We present quasi-simultaneous, multi-epoch radio and X-ray measurements of Holmberg~II~X-1 using the European VLBI Network (EVN), the Karl G. Jansky Very Large Array (VLA), and the {\it Chandra} and {\it Swift} X-ray telescopes. The X-ray data show apparently hard spectra with steady X-ray luminosities 4 months apart from each other. In the high-resolution EVN radio observations, we have detected an extended milli-arcsecond scale source with unboosted radio emission. The source emits non-thermal, likely optically thin synchrotron emission and its morphology is consistent with a jet ejection. The 9-GHz VLA data show an arcsecond-scale triple structure of Holmberg~II~X-1 similar to that seen at lower frequencies. However, we find that the central ejection has faded by at least a factor of 7.3 over 1.5 years. We estimate the dynamical age of the ejection to be higher than 2.1 years. We show that such a rapid cooling can be explained with simple adiabatic expansion losses. These properties of Holmberg~II~X-1 imply that ULX radio bubbles may be inflated by ejecta instead of self-absorbed compact jets. 
\end{abstract}

\begin{keywords}
accretion, accretion discs -- black hole physics -- X-rays: binaries
\end{keywords}

\section{Introduction}

Ultraluminous X-ray sources (ULXs) host accreting compact objects, whose luminosity ($L_X>3\times 10^{39}$~erg/s) exceeds the Eddington limit of a 20-M$_{\odot}$ black hole (BH). ULXs are a mixed bag of sources. Some may host super-Eddington neutron stars \citep{Bachetti:2014kx} or stellar-mass ($M=3-20$~M$_{\odot}$) BHs \citep{Liu:2013fk,Motch:2014uq}. Some of them are likely to be intermediate-mass  $M=10^{2}-10^{5}$~M$_{\odot}$ BHs \citep[][]{Farrell:2009ys,Feng:2010fk,Pasham:2014vn,Mezcua:2015uq} and may also host recoiling super-massive BHs \citep{Jonker:2010ys}. 

Given the tentative observational evidence for the anti-correlation between BH mass and host galaxy metallicity \citep{Crowther:2010fk} and that some ULX progenitor stars could potentially have been massive enough \citep[e.g.][]{Grise:2012li,Poutanen:2013cr} to collapse directly to a BH, ULXs might host massive stellar-mass (20-100~M$_{\odot}$) BHs  \citep{Fryer:1999fk,Mapelli:2009fk,Zampieri:2009if,Belczynski:2010kx,Mapelli:2013uq,Bachetti:2013zr}. In fact, Holmberg~II~X-1 lies in a relatively metal-poor dwarf irregular galaxy \citep{Prestwich:2013bh,Brorby:2015kx}. Such a mass range has been shown to fit its observed X-ray and radio properties \citep{Goad:2006fk,Cseh:2014fk} and may also be valid for IC~342~X-1 \citep{Marlowe:2014fk}.

To date, relatively few ULX radio counterparts have been studied, and these few show a relatively diverse nature. The first radio bubble, associated with NGC~5408~X-1, was discovered about a decade ago \citep{Kaaret:2003pz} although one other, associated with NGC~6946~X-1, was detected earlier but considered as a super- or hypernova remnant \citep{van-Dyk:1994kc}. Following the discoveries of more radio bubbles \citep[e.g.][and references therein]{Cseh:2012fk}, transient jets were also caught in two non-persistent ULXs, which in both cases were associated with X-ray outbursts of the central compact object \citep{Webb:2012kx,Middleton:2013uq}. Furthermore, in a few other cases, apparently steady radio emission has been detected on milli-arcsecond scales \citep{Mezcua:2011lp,Mezcua:2013kx,Mezcua:2015uq} and interpreted as self-absorbed radio cores, which are typically seen during radiatively inefficient accretion states of Galactic BH X-ray binaries \citep[e.g.][]{Fender:2009bh}. 

Focusing on the large-scale environments of ULXs, jets have been argued to be a plausible supply mechanism responsible for inflating the observed radio bubbles (and shock-ionized optical bubbles) \citep[e.g.][]{Cseh:2012fk}. On the other hand, direct evidence of jets in Holmberg~II~X-1 \citep{Cseh:2014fk} has been found, despite an environment that appears to be dominated by photoionisation \citep{Lehmann:2005bu,Egorov:2013ys}. This may indicate that shock ionization is not a necessary consequence of jets. Shock-ionized optical bubbles \citep[e.g.][]{Pakull:2002fk,Pakull:2008dd} could equally be inflated by powerful winds. These winds are thought to originate from a supercritical disk that may also give rise to geometrically collimated X-rays \citep{Poutanen:2007vn}, although calorimetry of photoionized optical bubbles argues against strongly beamed X-ray emission \citep[e.g.][]{Pakull:2002fk,Kaaret:2009uq}. Moreover, direct observational evidence for super-Eddington winds in ULXs is scarce \citep{Walton:2012bh,Walton:2013qf,Middleton:2014kx}. Interestingly, a similar group of objects, such as S26 in NGC~7793 and MQ1 in M83, which do not appear to have luminous central X-ray sources, can nonetheless have more energetic optical bubbles \citep{Pakull:2010kq,Soria:2014ve} than classical ULXs. These two peculiar examples both appear to have radio jets (or at least hot spots) and relatively large cocoons. Such morphology resembles SS433, although the total energy content of the W50 bubble surrounding SS433 is two orders of magnitude lower \citep{Pakull:2010kq}.

In the Galactic BH X-ray binaries, there is an observational anticorrelation between the presence of disk winds and jets. Winds are observed in soft or disk-dominated objects and jets are typically seen (at lower Eddington rates) in hard-state objects \citep[e.g.][]{Ponti:2012kx}. Broadly speaking, in active galactic nuclei (AGNs), near-Eddington luminosities are also coupled with powerful winds, whereas relativistic jets tend to occur at lower accretion rates \citep[e.g.][]{King:2015uq}. The presence of a collimated jet structure in Holmberg~II~X-1 coupled with a high X-ray luminosity might be somewhat extraordinary if one assumes persistent near- or super-Eddington accretion. However, little is known about switching between sub- and super-Eddington accretion. Possibly the best examples for such transitions are the Galactic systems GRO~J1655--40 \citep[e.g.][]{Meier:1996fk} and GRS1915+105 \citep[e.g.][]{Arai:2009uq}. Additionally, the higher BH masses inferred for sources like Holmberg~II~X-1 may imply longer characteristic timescales \citep{McHardy:2006zk}.

In summary, Holmberg~II~X-1 is a bona fide ULX with a typical X-ray luminosity of $\sim10^{40}$~erg~s$^{-1}$ \citep[e.g.][]{Grise:2010jf}. It is hosted by a nearby, relatively metal-poor \citep[e.g.][]{Prestwich:2013bh} dwarf irregular galaxy at a distance of 3.39~Mpc \citep{Karachentsev:2002fk}. It is associated with an optical \citep[e.g.][]{Kaaret:2004ta} and a radio bubble \citep{Miller:2005xh}. The optical environment is photoionized with an average isotropic X-ray luminosity of $4\times10^{39}$~erg~s$^{-1}$\citep{Kaaret:2004ta}. The radio bubble is inflated due to jet activity and the deduced average jet power of $Q_j\sim2\times10^{39}$~erg~s$^{-1}$ \citep{Cseh:2014fk} is comparable to the average X-ray luminosity. Its X-ray spectrum shows typical ULX features, like a soft excess and a spectral break \citep{Kajava:2012fk}. Based on the energetics, the X-ray spectral properties, and the environment, its mass is likely to be in the range of 25-100~M$_{\odot}$ \citep{Goad:2006fk,Cseh:2014fk} with an accretion rate near or above the Eddington-limit. The high accretion rate may also be supported by that the jet morphology indicates a recurrent activity (see below). Also, the above mass range might be supported by the longer activity of the source as compared to Galactic microquasars \citep{Cseh:2014fk}.

How these massive stellar-mass BHs evolve remains unclear. Early BH growth is often hypothesised to proceed via (intermittent) super-Eddington accretion \citep[e.g.][]{Volonteri:2014dq,Pacucci:2015fk}. To investigate if some ULXs and their observable properties can act as a proxy for super-Eddington BH growth in the early Universe, we present quasi-simultaneous and multi-epoch observations of the ULX Holmberg~II~X-1  (hereafter Ho~II~X-1), which are summarised in Table \ref{obs}. Our primary aim is to study the nature and evolution of the previously-detected central radio source \citep{Cseh:2014fk}, and to investigate the link between the radio emission and the X-ray behaviour. In Sec. \ref{oar} we describe our observational results. In Sec. \ref{disc}, we first discuss the X-ray spectra and the high-resolution radio properties of Ho~II~X-1, followed by our interpretation of the temporal evolution of the radio ejecta in Ho~II~X-1. Finally, we discuss the physical properties of the jet ejecta.

\section{Observations, Analysis, and Results}\label{oar}

\begin{table}
\caption{Summary of the observations\label{obs}}
\begin{tabular}{lcccc}
\hline\hline
Inst. & Mode &Frequency or& On-source&Observation\\
&or config.&energy band&time&date\\
\hline
EVN & -- &  1.6 GHz & 12.5 h & 14 Jan 14\\
Swift &  PC &  0.3 -- 10 keV & 20.5 ks& 17--22 Jan 14\\
\hline
EVN & -- &  5 GHz & 12.5 h & 28 Apr 14\\
Chandra& 1/8 & 0.3 -- 10 keV& 11.8 ks& 25 May 14\\ 
VLA & A & 8 -- 10 GHz & 171 min& 25 May 14\\
\end{tabular}
\end{table}

\subsection{Joint {\it Swift}/XRT and EVN observations}

\subsubsection{{\it Swift}/XRT observations}\label{swift}

The {\it Swift} X-ray telescope (XRT) observed Ho~II~X-1 (PI: Gris\'e) in photon counting (PC) mode on 17, 20, and 22 Jan 2014 starting from 13:36:59 UT, 00:48:59 UT, and 02:27:58 UT, for useful exposures of 5.71 ks, 5.88 ks, and 8.96 ks, respectively. We retrieved level 2 event files that were subjected to standard screening methods\footnote{http://heasarc.nasa.gov/docs/swift/analysis/}. The source spectrum was extracted using an aperture of 20 pixels, which corresponds to 90\% of the PSF at 1.5 keV.  The background was extracted using an aperture of radius 40 pixels, centred on a J2000 location of RA=08:19:51.437 and Dec=+70:47:46.29. The auxiliary response file (ARF) was created with {\sc xrtmkarf}, which accounts for PSF correction. 

After rebinning the spectrum to have a minimum of 30 counts per bin, we fit the X-ray spectrum using {\sc XSPEC} v12.8.2 \citep{Arnaud:1996cr} and a response matrix of swxpc0to12s6\_20130101v014.rmf\footnote{http://heasarc.gsfc.nasa.gov/docs/heasarc/caldb/\-data/swift/xrt/index.html}. We fit the spectrum with an absorbed power law model ({\sc tbabs*tbabs*powerlaw}) and froze the first hydrogen absorption column density to the Galactic value of $N_H = 3.4 \times 10^{20}$~cm$^{-2}$ \citep{Dickey:1990uq}. We found a good fit of $\chi^{2}$/DoF= 0.99 for 80 degrees of freedom (DoF), a photon index of $\Gamma=1.91\pm0.08$, an equivalent hydrogen absorption column density of $N_H = (5\pm2) \times 10^{20}$~cm$^{-2}$, and an unabsorbed flux of $(6.6\pm0.4)\times10^{-12}$~erg~cm$^{-2}$~s$^{-1}$ in the 0.3--10 keV energy range. The fitted power-law photon index of $\Gamma=1.91$ is consistent with a hard X-ray spectrum. The flux corresponds to a luminosity of $(9.1\pm0.6)\times 10^{39}$~erg~s$^{-1}$ at the source distance of 3.39\,Mpc \citep{Karachentsev:2002fk}. We fit other models to the X-ray data. First a disk blackbody model, that gave a poor fit with a reduced $\chi^{2}>3$. A cutoff power law model gave an adequate fit, but with a cutoff energy at 500 keV, indicating that a simple power law fits well. A combination of a disk blackbody with a power law resulted a 0 keV inner disk temperature, indicating that a simple power law fits well. 

Motivated by the best-fit model of \citet{Walton:2015fk}, we also fit the {\em Swift} spectrum with two disc blackbody models ({\sc tbabs*tbabs*(diskbb+diskbb)}). This fit resulted in a reduced $\chi^{2}$/DoF=0.94, that may indicate an overconstrained model. The first hydrogen absorption column density was frozen to the Galactic value. The fit did not require any additional $N_H$ on top of the Galactic one. The fit resulted in inner disc temperatures of $T_{in,1}=0.32\pm0.05$~keV and $T_{in,2}=1.7^{+0.3}_{-0.2}$~keV. The hotter component agrees within the errors with the result of \citet{Walton:2015fk} and the cooler component is inconsistent within the errors; it is somewhat hotter than found by \citet{Walton:2015fk}.

\subsubsection{VLBI observations}\label{vlbi}

\begin{figure}
\resizebox{3.3in}{!}{
\includegraphics{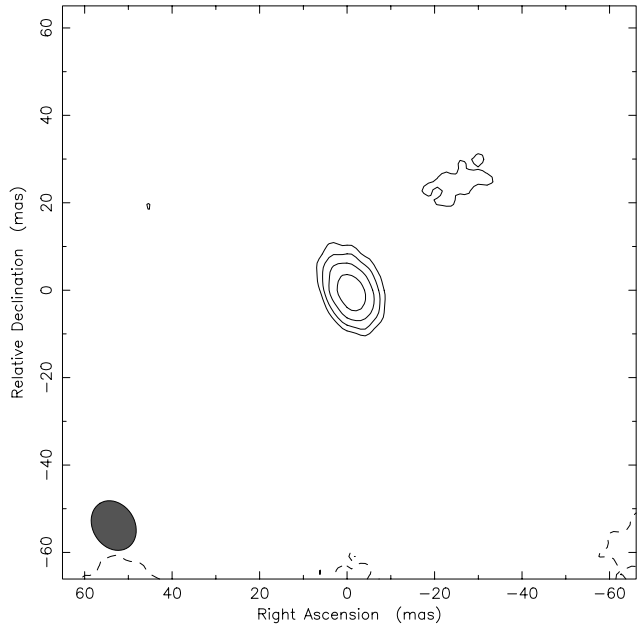}}
\caption{The 1.6-GHz e-EVN image of Ho~II~X-1 that was made using natural weighting. The lowest contours start at three times the rms noise level, at $\pm$15~$\mu$Jy beam$^{-1}$. Contours are increased by a factor of $\sqrt{2}$. The peak brightness is 53~$\mu$Jy beam$^{-1}$. The Gaussian restoring beam size is 12 mas $\times$ 9.6 mas at a major axis position angle of 32$^\circ$.}
\label{evn}
\end{figure}

We carried out very long baseline interferometry (VLBI) observations of the central component of Ho~II~X-1 with the electronic European VLBI Network (e-EVN) under program codes EC045 and RC001 (PI: Cseh), at 1.6 GHz and 5 GHz, respectively. The 1.6-GHz observation started on 14 Jan 2014 15:01:10 UT and had a total duration of about 18 h. The 5-GHz observation started on 28 Apr 2014 13:01:10 UT and lasted for about 18 h. At 1.6 GHz, the participating stations were Effelsberg (Ef; Germany), Jodrell Bank Lovell Telescope (Jb; United Kingdom), Medicina (Mc; Italy), Noto (Nt; Italy), Onsala (On; Sweden), Sheshan (Sh; China), Toru\'n (Tr; Poland), and Westerbork (Wb; The Netherlands). For the 5-GHz observation, the stations were Ef, Jb Mark II, Nt, On, Tr, Wb, Sh, and Yebes (Ys; Spain). The data recording rate was 1 Gbit s$^{-1}$ per station (except Mc where it was 512 Mbit s$^{-1}$) and no data were received from Ys. There were eight subbands, each with 16 MHz bandwidth in both left and right circular polarisation.

The observations were carried out in standard phase-referencing mode, using J0841+7053 as the phase--calibrator. The observations cycled continuously between the target and the reference source, spending 3.5 min on the target in every 5-min block. The data were calibrated in {\sc AIPS} v31Dec12 \citep[e.g.][]{Greisen:2003jb} following standard procedures \citep{Diamond:1995et}, and imaging was performed using {\sc Difmap} \citep{Shepherd:1994zv}. Antenna-based gain correction factors were obtained from the phase calibrator after a global amplitude self-calibration step, which were fed back to AIPS and applied to the data. These corrections were typically less than 10\% at 1.6 GHz and less than 20\% at 5 GHz. Fringe-fitting was then repeated taking into account the clean component model of the calibrator.

At 5 GHz, we reached an rms noise level of 12~$\mu$Jy beam$^{-1}$, the Gaussian restoring beam size was 2.7 mas $\times$2.1 mas, and we did not detect Ho~II~X-1. We set a 3-$\sigma$ upper limit of 36~$\mu$Jy on the flux density of the source.

At 1.6 GHz, the rms noise level was $\sim$5~$\mu$Jy beam$^{-1}$ and the Gaussian restoring beam size was 12~mas $\times$ 9.6~mas in PA 32$^{\circ}$. Ho~II~X-1 was detected at a signal-to-noise ratio of $>10$ (Fig. \ref{evn}), with a peak brightness of 53~$\mu$Jy at a position of (J2000) RA=08:19:28.9835 and Dec=+70:42:18.9966 (accurate to within 1 mas). To check the compactness of the source we fit the brightness distribution with a Gaussian in both the image and the visibility plane. Using the AIPS task {\sc JMFIT} for the image-plane fit, we found a fitted peak brightness of $49\pm6$~$\mu$Jy beam$^{-1}$ and an integrated flux density of $84\pm14$~$\mu$Jy. The integrated flux density is considerably higher than the peak brightness, likely indicating a resolved source. 

A fit in the visibility plane using a point-source model in {\sc Difmap} resulted in a total flux density\footnote{The error on this and the subsequent flux densities includes a 5\% systematic uncertainty.} of $S_{\nu}=65\pm6$~$\mu$Jy. If we fit the source with a circular Gaussian model, we find a total flux density of $S_{\nu}=88\pm11$~$\mu$Jy and a full width at half maximum (FWHM) size of $16.4\pm1.7$~mas. A comparison of these flux densities also indicates that the source is likely resolved. The obtained size of $16.4\pm1.7$~mas exceeds the restoring beam size of 12 $\times$ 9.6~mas. Furthermore, we estimate a resolution limit \citep{Kovalev:2005fk} of $\sim$3~mas that is well sampled by the image cell size of 1~mas. 

To test at what angular resolution the source starts to be resolved, we produced an image with a restricted ($u,v$) range of 0--4~M$\lambda$.  The resulting image had a beam size of 33 $\times$ 30~mas. The source had a peak brightness of 67~$\mu$Jy beam$^{-1}$, and a fit to a circular Gaussian model gave a total flux density of 95$\pm$10~$\mu$Jy with a FWHM of $20\pm1.6$~mas. There is an apparent increase in all of the fit parameters, although they are consistent with the above values within the errors. We conclude that the source is resolved on VLBI scales. We set a lower limit to its angular size of 16.4$\pm$1.7~mas that corresponds to a projected size of $0.26\pm0.03$~pc at a distance of Ho~II~X-1. A short summary is shown in Table \ref{fit}. Should the true size be larger than 16.4~mas, we estimate an upper limit of $\sim60$~mas from our highest resolution VLA images.

\begin{table*}
\caption{Summary of results on the central component \label{fit}}
\begin{tabular}{ccccc}
\hline\hline
Instrument&Band&Flux density&$\theta$&$T_{\rm{b}}$\\
&&[$\mu$Jy]&[mas]&[K]\\
\hline
EVN&1.6 GHz&88$\pm$11&16.4$\pm$1.7&1.5$\times$10$^5$\\
EVN&5 GHz&$<$36 &--&--\\
VLA& 8 -- 10 GHz&12$\pm$5&$<$235&--\\
\hline\hline
Instrument&Band&Unabsorbed flux&$\Gamma$&$N_H$\\
&&[erg cm$^{-2}$ s$^{-1}$]&&[cm$^{-2}$]\\
\hline
Swift&0.3--10 keV&(6.6$\pm$0.4)$\times10^{-12}$&1.91$\pm$0.08&(5$\pm$2)$\times10^{20}$\\
Chandra&0.3--10 keV&(6.5$\pm$0.6)$\times10^{-12}$&1.88$\pm$0.13&(5$\pm$2)$\times10^{20}$\\
\end{tabular}
\medskip

The results of the Gaussian fits to the central radio component and the fit of an absorbed power-law model to the X-ray source. For the 1.6-GHz VLBI data, the brightness temperature ($T_{\rm{b}}$) value is also provided.
\end{table*}

\subsection{Joint Chandra and VLA observations}
\subsubsection{{\it Chandra} observations}
We observed Ho~II~X-1 with the {\em Chandra X-ray Observatory} under ObsID 15771 (PI: Jonker) using the Advanced CCD Imaging Spectroscopic array (ACIS) S3 chip in FAINT mode and with 1/8 subarray.
The observation was carried out on 25 May 2014, started at 03:21:52 UT, and had a useful exposure of 11.85 ks. There were no strong background flares. The data were processed in a standard way and we used the CIAO version 4.6 and the CALDB version 4.6.1 to further process the level 2 event files. 

First, we checked for background flares following standard procedures\footnote{http://cxc.harvard.edu/ciao/threads/filter\_ltcrv/}. We removed strong sources and then checked the extracted background light curve and we found no strong flares during the observation. However, Ho~II~X-1 is a relatively bright X-ray source that caused a readout streak, which we therefore removed as described in the {\it Chandra} science thread pages\footnote{http://cxc.harvard.edu/ciao/threads/acisreadcorr/}. Using {\sc wavdetect}, we detected Ho~II~X-1 with a total of 6406 photons, and extracted its spectrum in the 0.3--10 keV range using {\sc specextract}, with a circular aperture of radius 3 arcsec. 

Given that we employed 1/8 subarray, providing a frame time of 0.4 sec, the expected\footnote{http://cxc.harvard.edu/toolkit/pimms.jsp} pileup fraction of Ho~II~X-1 is $\sim$5\%. To fit the spectrum we used the {\it Sherpa} package and its built-in pile-up model, following the {\it Sherpa} thread pages\footnote{http://cxc.harvard.edu/sherpa/threads/pileup/index.html}. First, we grouped our spectrum to have 30 counts per bin and then subtracted the background. We used an absorbed power-law model {\sc (xstbabs*xstbabs*xspowerlaw)}, setting the initial value of the hydrogen absorption column density to the Galactic one of $N_H = 3.4 \times 10^{20}$~cm$^{-2}$. We used the 'Neldermead' optimisation method to find a global solution that resulted in a reduced $\chi^{2}$/DoF of 1.14, a best-fit value of the photon index of $\Gamma=1.88\pm0.13$, and a hydrogen absorption column density of $N_H = (5\pm2) \times 10^{20}$~cm$^{-2}$. The resulting pile-up fraction was $\sim$6 \%, in agreement with our estimate above. 

To determine the flux we fit the spectra without a pile-up model and froze the power-law index to $\Gamma=1.88$. We found an unabsorbed flux of $(6.5\pm0.6)\times10^{-12}$~erg~cm$^{-2}$~s$^{-1}$ in the 0.3--10 keV range, corresponding to $L_X=(9.0\pm0.8)\times 10^{39}$~erg~s$^{-1}$ at the source distance of 3.39 Mpc. We also attempted to fit the same X-ray models to the {\it Chandra} spectrum as in Sec. \ref{swift}. The spectrum is well fit by a power-law and does not require any additional component. The power law fit parameters and the X-ray luminosity are consistent with being the same for the {\it Swift} and the {\it Chandra} observations, which were separated by $\sim$4 months.

Following \citet{Walton:2015fk}, we fit the {\em Chandra} spectrum with two disc blackbody models in XSPEC ({\sc pileup*tbabs*tbabs*(diskbb+diskbb)}). We also fit a disc blackbody plus a p-free disk blackbody model ({\sc pileup*tbabs*tbabs*(diskbb+diskpbb)}). These fits resulted in a reduced $\chi^{2}$/DoF= 1.10 and  $\chi^{2}$/DoF= 1.09, respectively. These reduced $\chi^{2}$ values may indicate that employing two disks instead of a power law does not significantly improve the fit. The first hydrogen absorption column densities were frozen to the Galactic value. In both of the cases, the fit did not require any additional $N_H$ on top of the Galactic one. The first fit resulted in inner disc temperatures of $T_{in,1}=0.31\pm0.04$~keV and $T_{in,2}=1.6^{+0.3}_{-0.2}$~keV. The second fit resulted in $T_{in,1}=0.28\pm0.07$~keV, $T_{in,2}=2.1^{+1.1}_{-0.7}$~keV, and $p=0.6^{+0.3}_{-0.1}$. The corresponding fitted temperatures between the first and the second fit agree within the errors. Furthermore, given that the value of $p$ can be consistent with 0.75, with the standard disc, we cannot differentiate between a standard or a p-free disc model. 

The fitted parameters of the first fit are consistent with being the same for the {\it Swift} and the {\it Chandra} observations. However, the cooler component is somewhat hotter than found by \citet{Walton:2015fk}. On the other hand, if a disk blackbody plus p-free disk is employed, both, the cooler and hotter component agrees within the errors with the result of \citet{Walton:2015fk}.

\subsubsection{{\it VLA} observations}\label{vlao}

\begin{figure}
\resizebox{3.2in}{!}{
\includegraphics{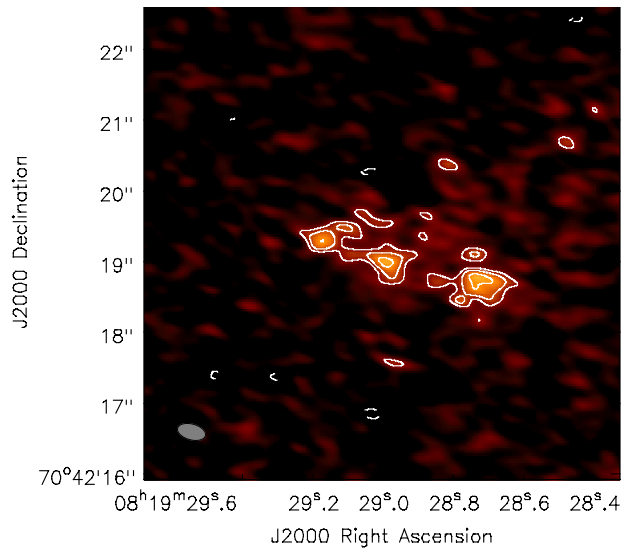}}
\caption{The VLA image of Ho~II~X-1 in X band, using Briggs robust=1 weighting. The lowest contours start at three times the rms noise level, at $\pm$6.6~$\mu$Jy beam$^{-1}$. Contours are increased by factors of $\sqrt{2}$. The peak brightness is 15~$\mu$Jy beam$^{-1}$. The Gaussian restoring beam size is 0.41" $\times$ 0.22" at a major axis position angle of 75$^\circ$ E of N. The background colour image is arbitrarily scaled.}
\label{Xband}
\end{figure}

The Karl G. Jansky Very Large Array (VLA) observed Ho~II~X-1 simultaneously with {\em Chandra} on 25 May 2014. The observations were carried out under project code SF0330 and ran from 03:44:50 UT for a total of 3.5 h. The VLA was in its most-extended A-configuration, and the correlator integration time was 2 s. Two 1-GHz wide basebands covered the 8--10 GHz range (X band) with a total of 16 spectral windows, each with 64 2-MHz wide channels. We calibrated the data using CASA \citep{McMullin:2007ys} version 4.1.0 and the VLA pipeline version 1.2.0. The absolute amplitude and bandpass calibrator was 3C286 and the gain and phase calibrator was J0721+7120. The primary calibrator was observed once for 6 min and the secondary calibrator was observed for 50 s in every 7.5-min block. The total on-source time was 171 min. There was no self-calibration applied. Images were made using Briggs robust=1 and robust=0 weighting and the multi-frequency synthesis (MFS) method with two Taylor terms. The rms noise was 2.2~$\mu$Jy/beam and the synthesised beam FWHM was 0.41${\prime\prime}$ $\times$ 0.22$^{\prime\prime}$ in PA 75$^{\circ}$ (Fig. \ref{Xband}). 

We detected three components, corresponding to those seen in C-band (4.5--6.5 GHz) VLA images \citep{Cseh:2014fk}. The central, NE, and SW components were detected with a signal to noise ratio of 6.6, 6.0, and 6.8, respectively. The in-band spectral index of the components could not be constrained at these signal-to-noise ratios. We modelled these components with elliptical Gaussians using the {\sc JMFIT} task within AIPS, and the results are shown in Table \ref{VLAfit}. We estimate the spectral index of the outer components between 5.24 and 9 GHz to be $\alpha_{\rm{SW}}=-0.5\pm0.4$ and $\alpha_{\rm{NE}}=-0.8\pm0.4$, where $S_{\nu}\propto\nu^{\alpha}$. However, we caution that the deconvolved size of the SW component is larger at X band than at C band. This could artificially flatten $\alpha_{\rm{SW}}$ and the size of the components at X band are poorly constrained (Table \ref{VLAfit}). 

For the central component, the size upper limits are in good agreement between the C-band uniformly-weighted image ($<0.28" \times 0.19"$) and the X-band robust 0 image (Table \ref{VLAfit}). Assuming a non-variable source, the spectral index of the central component is $\alpha_{\rm{C}}=-4.7\pm0.5$. We discuss possible interpretations for such a steep spectrum in Sec. \ref{temp}. Note that the minimum baseline length in the X-band data was 10.5~k$\lambda$. To test whether resolution effects could be artificially steepening the spectrum, we produced a C-band image with a minimum baseline length of 10.5k$\lambda$ and found no change in either the peak brightness or the integrated flux density. This indicates that the unresolved central component at C band is unlikely to be contaminated by additional extended emission that was resolved out in our X-band observations.

\begin{table*}
\caption{The VLA X-band components \label{VLAfit}}
\begin{tabular}{cccccl}
\hline\hline
Comp.&Peak&Flux dens.&Size&Rob-&PA\\
&[$\mu$Jy/b]&[$\mu$Jy]&[mas $\times$ mas]&ust&\\
\hline
Central&$12\pm3$&$12\pm5$&$<$ 230$\times$240 &0&76$^{\circ}$\\
Central&$13\pm2$&$50\pm10$&530$\pm$170$\times$480$\pm$190&1&62$^{\circ}$\\
SW&$14\pm2$&$64\pm11$&700$\pm$140$\times$450$\pm$110&1&91$^{\circ}$\\
NE&$12\pm2$&$35\pm7$&500$\pm$160$\times$300$\pm$200&1&94$^{\circ}$\\
\end{tabular}
\medskip

The Gaussian fit results of the different components.
\end{table*}

\section{Discussion}\label{disc}

\subsection{Is Ho~II~X-1 in the canonical X-ray hard state?}

It is common to fit ULX spectra with a two-component phenomenological model such as a power law and disk blackbody. On the other hand, high-quality {\em XMM-Newton} spectra of ULXs show a soft excess and a luminosity-dependent spectral break \citep{Stobbart:2006fk,Gladstone:2009fk}, and this is also the case for Ho~II~X-1 \citep{Kajava:2012fk}, that may indicate super-Eddington accretion. On the other hand, \citet{Caballero-Garcia:2010uq} suggested that such a spectrum might be reflection-dominated and that the dominant contribution to the total luminosity is coming from the power-law component, potentially implying the sources to be sub-Eddington. However, in the case of Ho~II~X-1, this model resulted in a fitted photon index that might be too steep for a canonical hard state \citep{Caballero-Garcia:2010uq}. Furthermore, {\em NuSTAR} observations have confirmed the presence of a high-energy cutoff for a number of ULXs \citep{Bachetti:2013zr,Walton:2014uq,Rana:2015fk}, which in those cases likely rules out a Compton hump arising from a reflection component in a canonical hard X-ray state. This has recently been shown specifically for Ho II X-1 \citep{Walton:2015fk}.

Thus, despite the fact that the current spectra are well fit by a single power law, we argue that Ho~II~X-1 is unlikely to be in the canonical hard state. Additionally, there is no hint of a self-absorbed compact jet in Ho~II~X-1 (see later), as is typically seen in this state. It is also unlikely that Ho~II~X-1 is in a non-standard, efficient hard state \citep{Rushton:2010jy,Coriat:2011qm} since the X-rays are steady and decorrelated from the radio emission. Also, our X-ray spectra may as well be fit with two disk blackbodies. In this case, it would be straightforward to exclude Ho~II~X-1 being in a canonical hard state, and such a fit possibly indicates a non-standard high state or a super-Eddington accretion state \citep{Walton:2015fk}.

In the following we investigate the temporal behaviour of the X-ray spectra. \citet{Grise:2010jf} presented multiple {\em Swift}/XRT observations of Ho~II~X-1, finding the source to remain in a soft spectral state for over 4 months in 2009/2010, in contrast to the hard spectra observed in 2006, which persisted for up to a month. The hard spectra from 2006 could also be fit with a disk blackbody and a power law model. This is contrary to our spectra where this combination of models resulted in a 0 keV inner disk temperature. On the other hand, the power-law components are identical within uncertainties between these {\em Swift} data sets, so a variable soft component is likely to be present. Also, comparing our {\em Swift} and {\em Chandra} spectra, the power-law components are consistent with being the same even though they are separated by 4 months. Based on our observations, we conclude that the (apparently) hard spectra are in general steady, as compared to the soft spectra, which, beyond a possibly variable soft component, are associated with irregular flares \citep{Grise:2010jf}. The potential link between these flares and the radio activity needs further investigation.

\subsection{VLBI morphology}

In light of the relatively stable X-ray spectra, in the following we shall investigate the contemporaneous behaviour of the central radio source that is coincident with the ULX position.

The 1.6-GHz VLBI component was found to be extended, with a size of $\sim$0.26~pc.  However, no milliarcsecond-scale radio emission was detected three months later at 5 GHz.  This non-detection at 5~GHz is consistent with the extended morphology found at 1.6 GHz, and also with the optically thin synchrotron spectrum observed at lower resolution \citep{Cseh:2014fk}. Thus, we infer that the VLBI emission is not due to a partially self-absorbed core jet. To estimate whether the source could be Doppler-boosted, we calculate its brightness temperature using:
\begin{equation}
\left( \frac{T_{\rm{b}}}{\rm{K}} \right)=1.22 \times 10^{6} (1+z) \left(\frac{S}{{\rm \mu Jy}}\right) \left( \frac{\rm{GHz}}{\nu}\right)^2 \left( \frac{\rm{mas}}{\theta} \right)^2
\end{equation}
After substituting $S_{\nu}=88$~$\mu$Jy at 1.66~GHz, $\theta=16.4$~mas, and $z=0.00079$, we find a brightness temperature of $T_{\rm{b}}\simeq1.5\times 10^5$~K. This is a relatively low value with respect to the equipartition value of relativistic compact jets of $T_{\rm{b,eq}}\simeq5\times10^{10}$~K \citep{Readhead:1994fk}. Given that the Doppler-factor is $\delta \simeq T_{\rm b} / T_{\rm{b,eq}}$, the low $T_{\rm b}$ of the emission argues against Doppler-boosted emission and hence against relativistic beaming being responsible for strong radio variability (see next section).

\subsection{Time-dependent radio behaviour}\label{temp}
 
Assuming a non-variable source, the spectral index of the central component between 5.24 and 9.0\,GHz is $\alpha_C=-4.7\pm0.5$, which is significantly steeper than the previous estimate of $\alpha=-0.8\pm0.2$ \citep{Cseh:2014fk}. In fact, such a steep spectrum is unphysical unless the synchrotron spectrum has a cutoff at the observed frequency due to, e.g., synchrotron ageing. We investigate this option in the following section.

Since the two VLA measurements were taken about 1.5\,yr apart, we therefore consider time variability. Taking the 5.24-GHz VLA peak brightness of 152~$\mu$Jy/beam and a spectral index of $\alpha=-0.8\pm0.2$, the expected peak brightness of a point source at 9 GHz is 99$\pm$11~$\mu$Jy/beam. The detected X-band peak brightness of $\sim$12~$\mu$Jy/beam is at least a factor of 7.3 lower than this (a difference of $\sim8\sigma$). Given that resolution effects were excluded (Sec. \ref{vlao}) and that the central source is point-like at both C and X band, this indicates that the flux density decreased with time.

\subsection{The jet properties of Ho~II~X-1}

The central component is due to optically thin synchrotron emission, consistent with a compact ejection. Since it was found to be a factor of $\sim$3 brighter than the outer components, this was interpreted as a renewed radio activity in the core \citep{Cseh:2014fk}. The current VLA observation, obtained 1.5 yr later, also shows the same morphology. However, the brightness of the central component is comparable to or less than that of the outer components. Therefore, the central component has cooled much faster than the outer components. This is consistent with the deduced spectral indices in Sec. \ref{vlao}. The central component has apparently faded by at least a factor of 7.3, is resolved on VLBI scales, and is not affected by Doppler boosting. 

The variation in the flux density over the course of $\sim$1.5 yr is likely a result of expansion. On the other hand, without a light curve, and with a VLBI component that is not (yet) resolved into a two-sided jet, the evolutionary stage of the ejection is unclear. Here, we assume a spreading jet and a Mach number similar to that of the large-scale jet \citep{Cseh:2014fk}. Then a lateral expansion velocity of $0.2c$ is expected for a speed of $c$ along the jet axis. This maximum expansion speed would result in a maximum radial increase of $\sim$4~mas (8~mas FWHM) over a year and may limit the age of the ejection to $\geq$2.1 years. On the other hand, the bulk speed is unknown and based on the geometry it could be as slow as $0.17c$. Considering such a mildly relativistic bulk speed, the radial expansion rate would be $\sim$1.3~mas/yr, implying an increased ejection age of $\sim$13--46~yr for our measured size of 16.4--60~mas.

In the following, we deduce an approximate adiabatic timescale ($t_{\rm{ad}}$) and compare it with the synchrotron cooling timescale ($\tau$). By assuming a fixed expansion speed of $v_{\rm{exp}}=0.2c$, the
adiabatic timescale can be approximated from the apparent physical size ($r$) of the ejection using $t_{\rm{ad}}\simeq\frac{3}{2}\frac{r}{v_{\rm{exp}}}$. For an inner structure of size 16.4--60~mas, this gives $t_{\rm{ad}}\simeq3.2$--$11.7$~yr. Next, we estimate the synchrotron lifetime of relativistic electrons at an observed frequency of 9 GHz, using $\tau\simeq2.693\times10^{13} \nu^{-1/2}B^{-3/2}$ \citep[e.g.][]{Longair:1994hs}, where the units are in Hz, mG, and yr. We estimate the magnetic field strength from minimum energy arguments, assuming equipartition between the energy in relativistic particles and that in the magnetic field. Using $B=1.8\times10^7\left(\frac{\eta L_{\nu}}{V}\right)^{2/7}\nu^{1/7}$~mG \citep[e.g.][]{Longair:1994hs}, and substituting in the VLBI parameters of $\nu=1.6\times 10^9$~Hz,  $V=3\times10^{47}$~m$^3$, and $L_{\nu}=1.21\times 10^{17}$~WHz$^{-1}$, we find $B\simeq0.77\eta^{2/7}$~mG, where $\eta-1$ is the ratio of energy in protons to that in relativistic electrons. Here we consider values of $\eta=1$--$100$, and find $\tau\simeq(1.8$--$13)\times 10^3$~yr\footnote{Here we do not perform the calculations with the size upper limit of $\sim$60~mas, because a higher size would mean even longer timescales.}. A comparison of these timescales indicates that the ejection is dominated by adiabatic losses, which results in the observed rapid flux decay. The outer components are bright because they are likely the working surfaces where the jet ejecta interact with the surroundings, allowing the conversion of kinetic energy to radiation. Finally, we note that with an average jet power of $Q_j\simeq2.1\times10^{39}$~erg~s$^{-1}$, the surrounding radio bubble might have been inflated in just $\sim$390~yr \citep{Cseh:2014fk}.

\subsection{Summary and implications}
We have conducted quasi-contemporaneous X-ray and radio observations of Ho~II~X-1. The X-ray observations show steady, apparently hard spectra with properties that are identical to the power-law component observed in spectra from 2006 \citep{Grise:2010jf}. By contrast, other observations in 2009/2010 have shown soft spectra and irregular flaring activity \citep{Grise:2010jf}. Given the steady, apparently hard emission, the X-ray flux variability may well arise purely from the soft component. The VLBI data reveal an extended radio source with a size of 0.26~pc. Based on its variability and brightness temperature, we find that it is consistent with non-thermal emission and that it is not affected by Doppler boosting. This is consistent with a scenario where the radio emission is due to a jet ejection event. Assuming a laterally-expanding jet, we estimate the dynamical age of the ejection to be $\geq$2.1~yr. Whether an ejection event could be correlated with the soft or flaring nature of the X-ray spectra remains to be investigated. On the other hand, we argue that the hard X-ray spectra are unlikely to correspond to a canonical hard state. This conclusion is also supported by the lack of a classical flat-spectrum, partially self-absorbed compact radio jet. Also, the lack of any correlation between the X-rays and the radio emission may also rule out radiatively efficient hard states, i.e. the `outlier branch' on the X-ray/radio correlation of BH X-ray binaries \citep[e.g.][]{Coriat:2011qm}. Using the VLA we have detected a triple structure at 8--10\,GHz that corresponds well to the morphology seen at lower frequencies. However, the central component has faded by at least a factor of 7.3 in just 1.5 years. We explain the observed cooling with adiabatic expansion losses that dominate over synchrotron cooling, regardless of the jet composition (baryonic or leptonic).

\citet{Cseh:2012fk} reported unresolved radio emission in the vicinity of the ULX IC~342~X-1 and concluded that it was inconsistent with a self-absorbed compact jet, and could instead be a clump of emission from the radio nebula. On the other hand, \citet{Marlowe:2014fk} argued that this unresolved emission could be consistent with an expanded ejection similar to that in Ho~II~X-1. By comparison, it is indeed plausible that such a radio component in IC342~X-1 could have faded on a timescale of a few years. If that is the case, then the quoted size limit on the ejection \citep{Marlowe:2014fk} of IC~342~X-1 might be too large, because the spatial variation would be small even with highly relativistic expansion speeds.

In light of our results on Ho~II~X-1, it is possible that other ULX radio bubbles could also have been inflated by transient ejecta. This could imply that it is not necessary to invoke self-absorbed compact jets to inflate ULX bubbles. Moreover, the self-absorbed compact jet in Cyg~X-1 has been argued to be incapable of inflating its bubble \citep{Heinz:2006uq} due to the mismatch between the energetics of the jet and the bubble. On the other hand, expanding ejecta and disk winds, which are potentially triggered by high accretion-rate events, could serve as a more plausible explanation in the above cases, as they might indeed be capable of carrying the required energy.

Ho~II~X-1 shows a collimated jet structure that is coupled with a high average X-ray luminosity whose ionizing rate is comparable to the average jet power. This may be unexpected given that AGNs with high accretion rates show strong winds but no powerful jets \citep[e.g.][]{Gan:2014fk,Tombesi:2015uq}. Similarly, in the context of regular BH X-ray binaries the jet is also suppressed in disc-dominated states, and \citet{Meier:1996fk} argues that a super-Eddington wind could be a secondary jet-suppression mechanism.  Nonetheless, Ho~II~X-1 does show discrete jet ejecta. This may imply that if the current X-ray luminosity is super-Eddington, then Ho~II~X-1 has to switch between sub- and super-Eddington rates to form jet ejecta. Alternatively, it could be that the source is not super-Eddington, but is rather in a persistent high, intermediate state where jet ejecta are thought to be released whenever crossing the so-called `jet line'. 

X-ray monitoring of the source with triggered radio observations could differentiate between any of the cases and test if X-ray flaring traces a jet ejection event.

\section*{Acknowledgments}
DC thanks S. Thoudam for useful discussions. The National Radio Astronomy Observatory is a facility of the National Science Foundation operated under cooperative agreement by Associated Universities, Inc. The scientific results reported in this article are based in part on observations made by the Chandra X-ray Observatory. The European VLBI Network is a joint facility of European, Chinese, South African, and other radio astronomy institutes funded by their national research councils.  JCAMJ is the recipient of an Australian Research Council Future Fellowship (FT140101082), and also acknowledges support from an Australian Research Council Discovery Grant (DP120102393). EK acknowledges fund from an NWO Vidi grant Nr. 2013/15390/EW. We acknowledge support from the Hungarian Scientific Research Fund (OTKA NN110333). SC acknowledges financial support from the UnivEarthS Labex program of Sorbonne Paris Cit\'e (ANR-10-LABX-0023 and ANR-11-IDEX-0005-02).

\bibliographystyle{mn2e} 

{\small \bibliography{my}}
\label{lastpage}

\end{document}